\documentclass[a4paper,fleqn,usenatbib]{mnras}

\usepackage[T1]{fontenc}
\usepackage{ae,aecompl}

\usepackage{graphicx}	
\usepackage{amsmath}	
\usepackage{amssymb}	
\usepackage{tabularx}
\usepackage{graphicx}
\usepackage{txfonts}
\usepackage{longtable}
\usepackage{hhline}
\usepackage{multirow}
\usepackage{lscape}
\usepackage{array}

\title[Snowline migration]{Effects of disc midplane evolution on CO snowline location}
\author[Pani\'c, O. \& Min, M.]{Pani\'c, O.$^{1,2}$\thanks{E-mail:o.panic@leeds.ac.uk}\thanks{Royal Society Dorothy Hodgkin Fellow} and Min, M.$^{3,4}$
\\ $^{1}$School of Physics and Astronomy, University of Leeds, Leeds LS2 9JT, United Kingdom
\\ $^{2}$Institute of Astronomy, Madingley Road, Cambridge, CB3 0HA, United Kingdom
\\ $^{3}$SRON Netherlands Institute for Space Research, Sorbonnelaan 2, 3584 CA Utrecht, The Netherlands
\\ $^{4}$Astronomical Institute Anton Pannekoek, University of Amsterdam, Science Park 904, 1098 XH Amsterdam, The Netherlands
}

\date{Accepted XX. Received YY; in original form ZZ}

\pubyear{2015}

\begin{document}
\label{firstpage}
\pagerange{\pageref{firstpage}--\pageref{lastpage}}
\maketitle

\begin{abstract}

Temperature changes in the planet forming disc midplanes carry important physico-chemical consequences, such as the effect on the locations of the condensation fronts of molecules - the snowlines. Snowlines impose major chemical gradients and possibly foster grain growth.
The aim of this paper is to understand how disc midplane temperature changes with gas and dust evolution, and identify trends that may influence planet formation or allow to constrain disc evolution observationally. We calculate disc temperature, hydrostatic equilibrium and dust settling in a mutually consistent way from a grid of disc models at different stages of gas loss, grain growth and hole opening. We find that the CO snowline location depends very strongly on disc properties. The CO snowline location migrates closer to the star for increasing degrees of gas dispersal and dust growth. Around a typical A type star, the snowline can be anywhere between several tens and a few hundred au, depending on the disc properties such as gas mass and grain size. In fact, gas loss is as efficient as dust evolution in settling discs, and flat discs may be gas-poor counterparts of flared discs. Our results, in the context of different pre-main sequence evolution of the luminosity in low- and intermediate-mass stars suggests very different thermal (and hence chemical) histories in these two types of discs. Discs of T Tauri stars settle and cool down while discs of Herbig Ae stars may remain rather warm throughout the pre-main sequence.  

\end{abstract}




\section{Introduction}


Understanding the physical and chemical structure of protoplanetary discs is crucial if we want to understand the diversity of the observed exosolar planets \citep{helled2014}. The dense midplane region of a protoplanetary disc is where planets form, interact with the disc and migrate. In particular, the outer disc regions are of interest for giant planet formation \citep[e.g., see review by][]{helled2013}.
Midplane is generally well hidden from the view of astronomical telescopes due to high optical depth, except for the millimetre wavelengths where the high sensitivity Atacama Large Millimetre-Submillimetre Array (ALMA) is able to detect both C$^{18}$O gas and dust emission from the cold outer disc midplane. Such detections remain few, with the brightest and best resolved disc HD163296 \citep[e.g.][]{rosenfeld2013,boneberg2016}.

Models of planet formation often rely on very basic assumptions on the disc structure, in which the temperature \citep[e.g.,][]{pollack1996,weidenschilling1997}, and sometimes also the density \citep[e.g.,]{ida2005,kenyon2006}, are constant with time. Most models focus on the effects of viscous heating but neglect the role of irradiative heating of the disc, dominant at larger orbital distances. Temperature structure of the midplane is important for planet formation because it determines the sound speed, which in turn influences the gravitational stability and radial drift of particles. Moreover, temperature-critical processes such as freeze-out and evaporation of molecular species ultimately determine the abundances of C, N and O in gas phase and in ice-mantles on grains, thereby affecting elemental abundances of the raw materials used to make rocky and gaseous planets \citep[e.g.,][]{helling2014,marboeuf2014}. It has been shown that accounting for the effects of thermal evolution of the disc can drastically change the outcome of planet migration models \citep{paardekooper2010}. 

Sophisticated calculations of disc heating help us to understand how the locations of condensation fronts of major molecular species such as H$_2$O and CO, known as snowlines, migrate. These locations are expected to host an enhancement in solids \citep{stevenson1988} and therefore enable more efficient particle growth leading to the formation of planets. The H$_2$O snowline has been in the spotlight of theoretical and observational efforts because of its relevance for terrestrial planet formation \citep[e.g.,][]{alessi2016}. In the recent years, the CO snowline is increasingly studied, and this is motivated both by the imaging of the CO snowline with the Atacama Large Millimetre/Submillimetre Array (ALMA) \citep{qi2011,mathews2013,qi2013a,qi2015}, and by the theoretical work finding that the chemical conditions specific to the CO snowline are consistent with the chemical compositions of Uranus and Neptune in our own Solar system \citep{ali-dib2014}. The region inward from the CO snowline, where temperature ranges between 20-40~K, is the region of the highest C/O gas-phase abundance in a disc \citep{oberg2011} and presents a potential birthsite of C-enriched gas giants \citep{madhusudhan2014}. Because the desorption of the CO ice is temperature-critical process, even relatively small changes to the temperature (e.g. 30$\%$) can cause major shifts in the CO snowline location and consequently, in the chemical composition of the midplane gas.

Heating of the midplane disc regions beyond the first several au from the star is provided mainly through the stellar radiation captured in the high disc layers and therefore it strongly depends on the vertical thickness of the disc. Traditionally, much of the work on disc structure relies on the spectral energy distribution (SED) analysis. This has led to an observational classification of discs 
where large excess emission in the mid- to far-infrared regime corresponds to vertically thick, flared discs, while the sources with relatively low excesses correspond to thinner, settled discs \citep[e.g.][]{meeus2001,vanboekel2005}. A natural explanation for disc settling is given by grain growth, whereby larger dust settles to the disc midplane, efficiency with which the dust disc captures the stellar light decreases, and the consequently lower temperature leads to lower excess emission in the `flat' SED type \citep{DD2004a,DD2004b}. For a decade, this has been widely regarded as the explanation behind the flared vs. settled SED dichotomy. Our work expands on this approach, by fully considering the hydrostatic aspect of the disc structure and the close interplay of gas and dust in the disc heating and settling. As outlined in the aforementioned papers, the gravitational pull a grain feels towards disc midplane is counterbalanced by the gas pressure, which in turn depends on the gas temperature. Gas temperature however is set by the dust temperature - value set by the thickness of the dust disc \citep[e.g.,][]{bruderer2012}. Therefore, we do not only consider the effects of dust evolution on disc settling (i.e., temperature) but also employ fully self-consistent modelling of the vertical structure with respect to heating, hydrostatic gas pressure and settling. A new aspect in this work is that we do not explore only dust-related processes but also the effects of gas loss on disc settling. This work is particularly timely, considering the ability of the newly constructed ALMA observatory to probe disc flaring \citep{degregorio2013} and CO snowline location \citep{qi2015}.

The primary aim of this paper is to quantify the extent to which gas and dust evolution alter the temperature conditions in disc midplane, and cause shifts in the CO snowline location. Specifically, we explore the effects of gas dispersal, grain growth and settling, loss of dust, inner clearing and viscous spreading. We keep the stellar parameters fixed, in order to be able to study the effects of disc properties in isolation. For the same reason, disc properties such as gas density, size of the cavity or dust size, are varied individually, to best isolate and compare their respective effects on the snowline location.
In Sect.~\ref{modelling} we describe the radiative transfer code we use to compute the disc structure, our fiducial model, and the explored parameter space. In Sect.~\ref{results} we present the resulting temperature structures, snowline locations, and how these depend on the individual model parameters. 
In Sect.~\ref{obs} we present the resulting SEDs and their variations, and discuss the observational implications for (sub)millimetre imaging of discs. We then discuss the physical implications of the identified trends on protoplanetary disc evolution and planet formation. In this Section we combine the individual effects of the varied parameters to discuss whether disc evolution as a whole causes net inward or outward shifts in CO snowline location. We also consider whether the different stellar evolution for low- and intermediate-mass stars contributes or opposes the identified trends. In Sect.~\ref{conclusions} we summarise our key findings.

\section{Disc structure calculation}\label{modelling}

\subsection{Calculation of the disc temperature}

Observational evidence exists for several aspects of disc ageing: inner regions are cleared of dust, grains grow to millimetre sizes, small dust is depleted, gas is photoevaporated or dispersed. All these processes are interdependent to some extent and it has proved challenging to constrain their relative timescales. Because of this, we do not base our analysis on one specific evolution model that includes all the aforementioned processes and relies upon assumptions on how their timescales compare. Instead, we study each process separately, but calculating a series of models where we change a single disc property, e.g. gas mass. This allows us to identify the processes that cause significant changes to the disc midplane temperature, and predict observable evolutionary trends.

There are two essential and closely interdependent aspects of disc structure: heating and pressure support, and these are dominated by the dust and gas, respectively. The gas pressure sets the vertical thickness for the dust, and it does so according to the temperature and its vertical gradient. At the same time, the vertical thickness of the dust sets the fraction of stellar radiation that the disc will absorb and which will go towards the heating of disc interior. Due to high densities and the fact that disc interior is optically thick to stellar light, gas and dust are thermally coupled and thus the hydrostatic equilibrium solution for the gas in disc models is based on the dust temperature. However, in order to be fully self consistent, a disc model must ensure both the heating/cooling, settling and hydrostatic equilibrium are verified simultaneously. There are a few radiative transfer codes that perform iterative structure calculations to ensure the temperature converges to a physically consistent solution, and we use one such code \citep[MCMax,][]{min2009}.

The bulk of the disc interior is heated through the process in which the dust absorbs and re-emits photons, thereby redistributing the stellar radiation which was initially captured by the dust in the disc surface. The physics of this process is the same across different spectral types and luminosities.
For this reason we only explore one set of stellar properties (a Herbig Ae star), and expect the qualitative results of our study to apply to discs around T Tauri stars as well.


\begin{figure}
 \includegraphics[width=8.5cm]{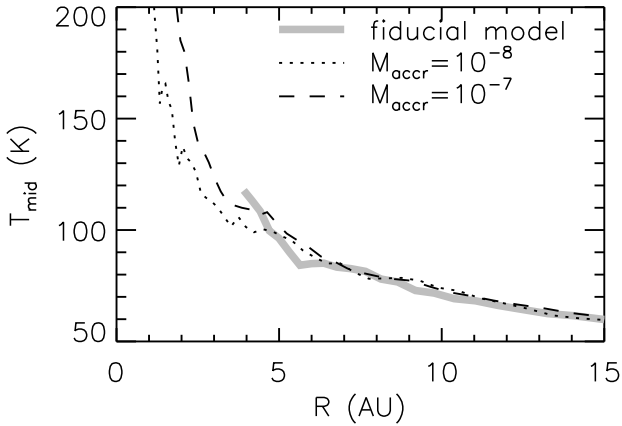}
 \caption{Temperature structure in the models with viscous heating (black) compared to our fiducial model with no viscous heating (grey). Mass accretion rates $M_{accr}$ ($M_{\odot}/yr$) are indicated in the legend. Highest density regions in the fiducial model require more than 5 million photon packages to reach convergence on the temperature structure - $T_{mid}$ at less then 4au is therefore omitted from the plot.}
\label{temp}
\end{figure}

Our disc models are heated by the radiation from the central star, and we do not include any external sources of radiation. In equilibrium, the interstellar radiation field typically heats dust to 10~K temperatures and therefore it dominates disc heating only in the cold and tenuous, outermost disc regions, well beyond the CO snowline. Discs exposed to significant external radiation may allow the non-thermal desorption of CO to take place in the low density regions at large radial distances from the star (e.g. the second CO snowline in IM~Lup disc \citet{oberg2015}).

In the innermost disc regions, viscous heating can be the dominant form of heating, particularly when the accretion through the disc is high. This effect, however, does not extend further than the first few au as we will show in Sect.~\ref{results} (Fig.~\ref{temp}), and concerns temperatures of roughly $\approx$100~K, and therefore the water snowline. We note that the extra heating in the first few au due to viscosity will have caused a vertically thicker disc locally, with possibly some shadowing, but this effect is not propagated further than 5~au, and we can see that the temperature has remained identical at large radial distances in all three models.
Because our work focuses primarily on the outer disc, we exclude viscosity as a source of heating in our calculations. Nonetheless, our results provide some insight in the behaviour of the H$_2$O snowline in the case when viscous heating is not dominant.

\subsection{The fiducial model and explored parameter space}\label{fiducial}

Our fiducial model is that of a disc around an 2~M$_{\mathrm{\odot}}$ A type star, with $T_{eff}$=10000~K and $L_{star}$=35~L$_{\odot}$, similar to HD~163296 \citep{vandenancker1998}. These properties are roughly matched by a 10~Myr old, 2.3~M$_{\odot}$, pre-main sequence star model in the database of \citet{siess2000}. 
Stellar and disc properties are held constant while we vary a single key disc structure parameter at a time as outlined in Tab.~\ref{tab1}. Our disc models have a radially continuous decrease in surface density following a power law of exponent $pow$, which we vary to both higher and lower values with respect to the fiducial value of 1. This power-law is consistent with observationally constrained values in \citet{andrews2007}, and is expected for the distribution of the bulk of the mass over a vast radial extent of viscously evolving discs (except for the exponential tail at the outer disc edge). The total disc mass is 1~$\%$ of the stellar mass, hence 2$\times$10$^{-2}$. We assume a global gas-to-dust ratio of $g/d$=100 \citep{reina1973,gorenstein1975}, the value found in the interstellar medium. In some model variations, the gas and dust mass are varied independently, effectively changing $g/d$. In all models, dust settling leads to dust overabundance closer to the midplane and therefore the global $g/d$ is not conserved locally, but there is a vertical dependence \citep[see also][]{mulders2012}.
The inner radius of the disc is set to the dust sublimation radius, roughly 0.24~au for a Herbig Ae star. As we will show, the exact value of the inner radius, at these scales, does not influence our results which concern larger radial distances. The outer radius is set to a 500~au from the star. 
The grain size distribution in our fiducial model has a slope of 3.5, and grains from $a_{min}=$0.01~$\mu$m, as in the interstellar medium, up to 1~mm in size as found in protoplanetary discs. 

The vertical extent of the gas disk and the settling of large grains towards the midplane of the disk are computed completely self-consistently using the iterative implementation described in \citet{mulders2012}. Vertical hydrostatic equilibrium of the disc is assumed. The settling efficiency of the grains depends on grain size, gas density and the turbulence parameter $\alpha$.

\textit{A note on turbulence.} A valid assumption for the regions of interest in this paper is that the dust scale height is proportional to $H_{dust} \propto H_{gas} \sqrt{\alpha/\tau_s}$ \citep{youdin2007}, where the stopping time $\tau_s\propto1/\rho_{gas}$ in the Epstein regime. Calculation of dust settling in MCMax uses this relation, so it is clear that the dust scale height has a direct dependence on the gas. In particular it depends on $\sqrt{\alpha\rho_{gas}}$. Therefore the turbulence parameter $\alpha$ and gas density $\rho_{gas}$ lend support to the vertical structure of the dust in the same way. Due to this direct parameter degeneracy, we fix the $\alpha_{turb}$ parameter (i.e., the $\alpha$ parameter from \citet{youdin2007}) to 10$^{-4}$ and keep this value constant. This is an order of magnitude lower than the upper limit on the turbulence inferred by \citet{flaherty2015} in the Herbig Ae disc around HD163296.



\begin{table}
 \centering
  \caption{Summary of the model parameters explored. The default values are underlined. Each of these parameters is varied while holding the remaining parameters fixed. In addition, gas and dust mass are also varied in tandem, while keeping their ratio at a constant value of 100.}
  \begin{tabular}{@{}lll@{}}
  \hline \hline
 Parameter &  & Explored values \\
\hline
$M_{dust}$ & Dust mass & 2$\times$ ( 10$^{-6}$, 10$^{-5}$, \underline{10$^{-4}$})~M$_{\odot}$ \\
$M_{gas}$ & Gas mass & 2$\times$ ( 10$^{-4}$, 10$^{-3}$, \underline{10$^{-2}$})~M$_{\odot}$ \\
$pow$ & Radial surface density slope & 0.8, \underline{1.0}, 1.2, 1.4, 1.6, 1.8 \\
$a_{min}$ & Minimum grain size & \underline{0.01}, 1, 100~$\mu$m \\
$a_{max}$ & Maximum grain size & \underline{1~mm}, 10~cm, 10~m, 1~km \\
$R_{in}$ & Size of the inner hole & \underline{0.24}, 2.5, 5.0, 7.5,...50~au \\
    \hline
\end{tabular}
\label{tab1}
\end{table}

The location of the CO snowline in our models is defined as the radial location in the disc midplane at the temperature of 20~K, as constrained with ALMA \citep{schwarz2016}. Assuming a different value of desorption temperature \citep[e.g.,][]{burke2010,noble2012} would consistently shift the snowline locations in all our models. However, this would have no bearing on the evolution trends that we identify. Because our models do not include an external radiation field, the temperature decreases monotonously with the radius and the CO snowline location is unique for each model. None of our disc models enter the regime in which the direct stellar radiation penetrates to the midplane. Irradiative heating by the central star dominates the disc midplane temperature in the region widely surrounding the CO snowline, and therefore our results on how this location changes with different disc parameters remain valid, regardless of the precise desorption temperature of CO ice. A single CO snowline radius (i.e., temperature) is a good approximation for ice desorption from the bulk of small, observable dust, and current imaging of CO snowline tracers (See Sect.~\ref{lines}) does not show evidence of the CO snowline being radially extended. If most of the ice is locked in large and fast-drifting particles we may expect to have a radially extended CO snowline due to the effect where desorption location is dependent on the dust/pebble particle size \citep{piso2015}.

\section{Results and discussion}\label{results}

\subsection{Effects on the CO snowline location}\label{snowline results}

\begin{figure*}
 \includegraphics[width=11.5cm]{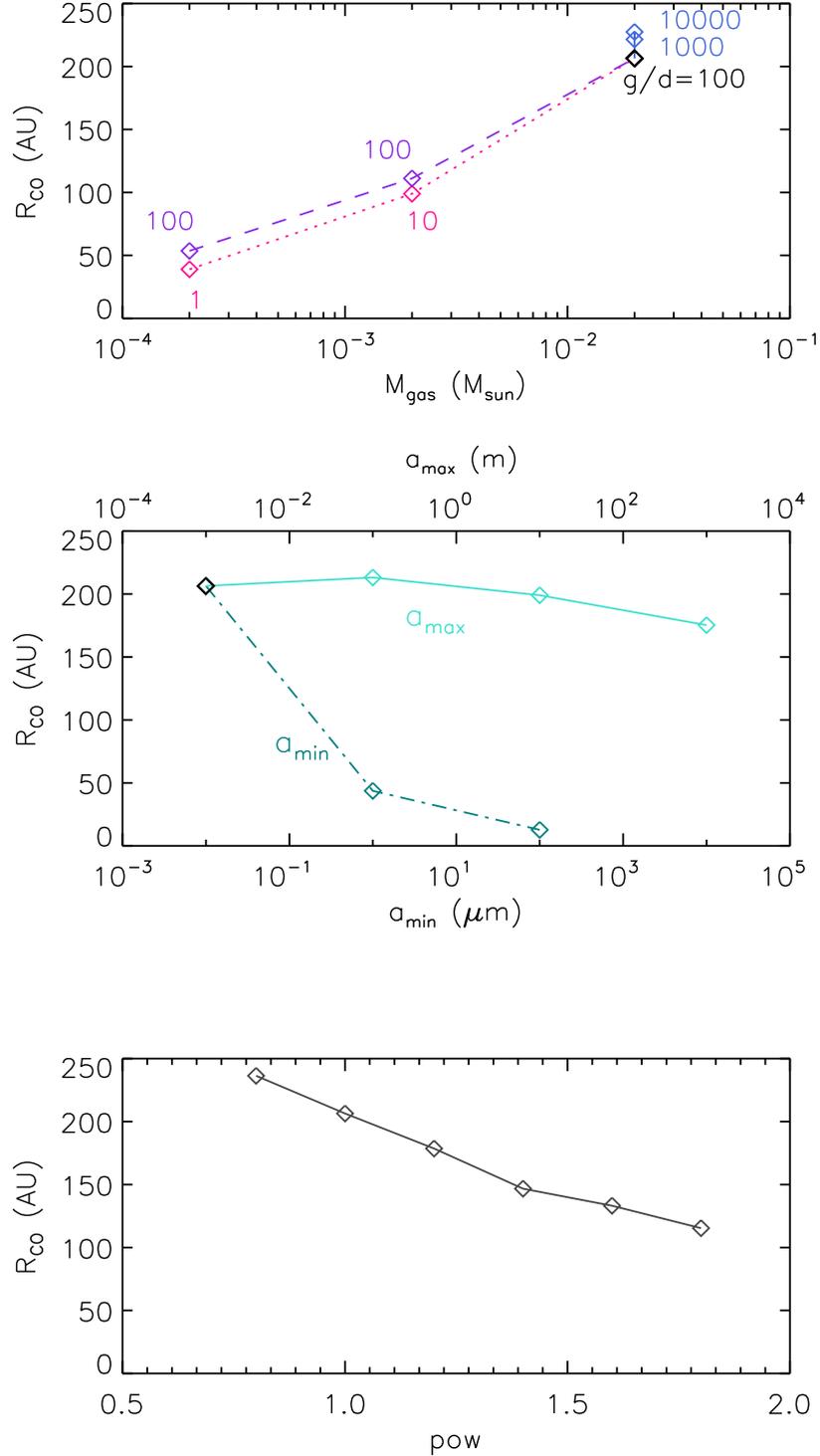}
 \caption{CO snowline location $R_{CO}$ for the different models. The fiducial model is shown in black. \textit{Top:} $R_{CO}$ at different gas masses, with additional variations of the dust mass (i.e., gas to dust mass ratio $g/d$), exploring different relative timescales for gas and dust loss. Purple dashed line illustrates a disc where the gas dissipates at a constant $g/d$ (i.e., at the same rate as the dust), pink dotted line a disc where only the gas is being removed while dust mass remains constant, and blue line a disc where only dust is being removed while the gas mass remains constant. \textit{Middle:} $R_{CO}$ for models where grain growth is simulated by increasing a$_{max}$ (m) (growth of large bodies in the disc midplane), shown with the full turquoise line, and models with increasing a$_{min}$ ($\mu$m) (growth of the smallest dust throughout the disc, including disc surface), shown with the dashed-dotted olive line.\textit{Bottom:} $R_{CO}$ for different surface density power-law distributions.}
\label{snowy}
\end{figure*}


In Fig.~\ref{snowy} we compare the locations of the CO snowline, $R_{CO}$, resulting from our parameter variations. By changing gas and dust masses in our model, we explore the effects of gas and dust loss - for example due to FUV photoevaporation and grain growth. It is important to reiterate that the timescale for dispersal of the gas and dust mass in the outer disc has not yet been constrained observationally. This concerns the \textit{total masses of discs}, i.e., disc material generally contained at 100~au distances from the star. We explore both the scenario in which the timescale for the loss of gas and dust is the same and the scenario in which one is lost on a much shorter timescale than the other \citep{takeuchi2005}. We explore disc dispersal by uniformly decreasing the column density throughout the disc. This is a valid assumption for almost the entire lifespan of the outer disc in different photoevaporation models \citep{alexander2006,clarke2007,owen2010}.

In the first scenario the canonical $g/d$=100 is preserved as gas and dust are lost at same rates. In Fig.~\ref{snowy} (upper panel) we can see a dramatic decrease in $R_{CO}$ while the masses of both gas and dust are reduced to 10$\%$ and then 1$\%$ of their initial values (dashed purple line). 
In these models, $R_{CO}$ decreases from roughly 200~au in the fiducial model to 100~au and then 50~au. 
We obtain a surprisingly similar, large shift in the CO snowline location, when only the gas mass is decreased (dotted pink line in Fig.~\ref{snowy}), indicating that the density of the gas, and not of the dust, plays a dominant role for indirect radiative heating of the outer disc. This is the first time that gas effects are measured separately from other disc parameters, and this approach allowed us to see that gas itself, is very important to irradiative heating of the outer disc. This fact is explained by the importance of gas density in providing the pressure needed to suspend the dust high above the midplane and in direct view of the star. A perhaps less intuitive result is that dust loss at a given gas mass results in a slight increase in the temperature and $R_{CO}$. In particular, we explore the scenario where only dust mass decreases, starting from our fiducial model at $\approx$70~M$_{\earth}$ to 7 and then 0.7~M$_{\earth}$ of dust, spanning the dust mass range observed from protoplanetary to debris discs \citep{panic2013}. Lower total dust masses have the effect of allowing starlight to penetrate slightly further into the disc surface and deposit energy at larger distances from the star due to the decrease in optical depth. At the same time, the fact that the small dust is still present in the disc, albeit with a lower density, ensures that the disc is efficiently heated. To summarise, the smallest dust grains, even when present in low numbers, are sufficient to provide the opacity in disc surface and keep the $\tau=$1 layer high. On the other hand, dissipation of the gas disc decreases the height of this layer and the disc heating significantly.

The effects of grain growth can be explored in a number of ways, and we chose the simplistic approach where we focus on grain growth in the two completely different regimes: 1) growth of the smallest dust grains in the disc surface whereby the small dust is completely removed, and 2) growth of the dust in disc midplane whereby large dust is produced. In these two scenarios we push the opacity and mass, respectively, to be provided by grains of larger sizes. The power-law of the dust size distribution is kept constant and, as in all our models, size-dependent dust settling is included. The results are shown in Fig.~\ref{snowy}, middle panel. To explore the first scenario, we increase $a_{min}$ from the ISM-like value of 0.01~$\mu$m to 100 and 1000 times larger sizes (dark green line). The effect on the $R_{CO}$ is dramatic as it moves from 200 to less than 50~au when $a_{min}$ increases to 1~$\mu$m. This large effect is understandable, because large grains settle closer to the midplane than small grains. Therefore the resulting $\tau=$1 surfaces in these models are lower and intercept less starlight, which leads to a temperature decrease. On the other hand, when $a_{max}$ is increased (cyan line), the snowline $R_{CO}$ shifts by comparably minor amounts. This result reflects the fact that the disc midplane, where the largest dust grains accumulate and grow, does not actively participate in the processes that lead to disc heating.

Radial distribution of disc material in a steady-state accretion disc follows the 1/$R$ dependence. However, we know from observations of millimetre dust emission that this power law may differ from one source to another \citep{andrews2007}, at least where millimetre grains are concerned. We therefore explore a range of power-laws in our models to investigate the effect that the radial distribution of material has on the midplane temperature and $R_{CO}$, from power-law dependence of $pow=$0.8 to 1.8, where our fiducial model follows the viscous $pow=$1.0. The results are shown in the lower panel of Fig.~\ref{snowy}. $R_{CO}$ is closer to the star for steeper radial density profiles, because larger density in the inner regions increases dust opacity. While the effect on the CO snowline is not as strong as in the case of gas removal and growth of the smallest dust grains, it is still worth considering when interpreting CO snowline observations in the context of an SED model, for example.

\begin{figure}
 \includegraphics[width=8.5cm]{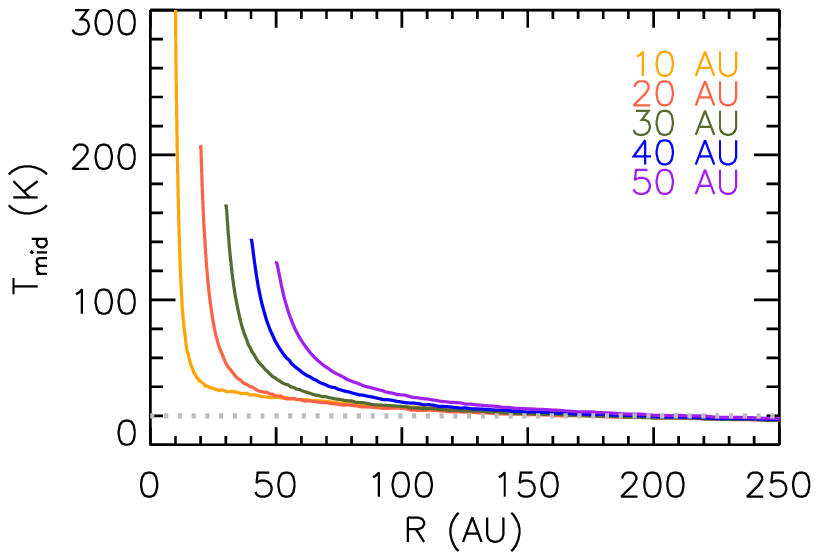}
 \caption{The changes in the midplane temperature structure as $R_{in}$ increases over the range indicated in the legend. The dashed grey line marks the 20~K CO desorption temperature. }
\label{rin}
\end{figure}

Disc dispersal driven by photoevaporation results in inside-out clearing of discs. We consider the effects of the complete removal of material, gas and dust, from the inner regions of the disc. We do this by moving the inner disc radius,  $R_{in}$, progressively to larger distances from the star in our models, up to $R_{in}=$50~au. The effect on the midplane temperature is significant only in the regions immediately behind the inner edge of the disc as shown in Fig.~\ref{rin}. The effect on the CO snowline location in these models are minor, as this location is considerably farther from the star. It is important to stress, however, that in the case of discs with lower stellar luminosity (e.g., around T Tauri stars) or more evolved gas and/or dust, the snowline can be close enough to the star to be affected by disc clearing. Any interpretations of the significance of the snowline location in such cases have to take into account the additional heating arising from the decreased optical depth in the inner regions.

The results obtained here provide a global description of trends in disc temperature. They show that the location of the CO snowline can change drastically depending on the total gas mass and on the size of the smallest grains in the disc surface. The CO snowline is therefore not a phenomenon that concerns a single radial location throughout the disc lifetime. The set of unique chemical and physical conditions associated with the snowline migrates through the planet forming regions in an evolving gas and dust disc. We will discuss the implications of these results on planet formation further in Sect.~\ref{planetformation}. Our approach simplifies the dependence of the CO desorption temperature on detailed properties such as the spatial structure of the grain and the degree of mixing of the CO and H$_2$O ice, by assuming that this temperature is 20~K. 
Regardless of the exact details of grain and ice structure in disc midplanes, our results show that the radial location where CO ice desorbs is strongly dependent on disc properties. 

\subsection{Effects on other snowlines}

\begin{figure*}
 \includegraphics[width=15cm]{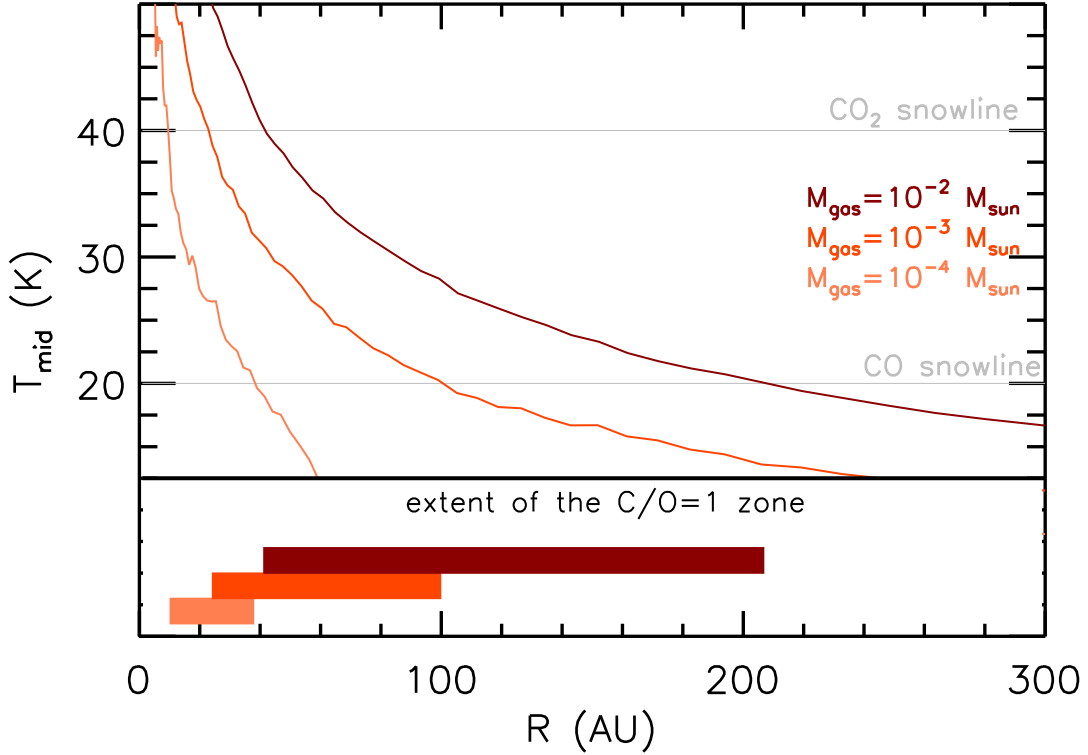}
  \caption{Midplane temperature profiles obtained for our fiducial model $M_{gas}=10^{-2}$~M$_{sun}$ and for the models with 10 and 100 times less gas mass.  Desorption temperatures adopted for CO and CO$_2$ are indicated with thin grey lines. The regions of maximum C/O ratio in the gas phase (C/O=1) between $R_{CO}$ and $R_{CO2}$ are indicated in the bottom panel for each of the models. }
\label{coratio}
\end{figure*}

As shown in Fig.~\ref{temp} the temperature of a vast extent of the disc midplane beyond the first 5~au is dominated by indirect radiative heating, encompassing not only the CO snowline but also snowlines of other chemically relevant species with desorption temperatures below 100~K. One such molecule is CO$_2$, an important carrier of O and C, and as such, a molecule whose abundance strongly influences the C/O ratio. Its desorption temperature is 40-50~K \citep{davis2005}, and the region delimited by the CO$_2$ and CO snowlines is the region of the maximum gas phase C/O abundance in the disc corresponding to C/O=1, where essentially all of the gas-phase C and O is contained in CO. This abundance is set mainly by the simple processes of adsorption/desorption of molecules, rather than a more indirect chemical pathway\citep{oberg2011}. One of the main results of our study is that the gas mass has a drastic influence on the heating of the disc midplane and hence on the location of these snowlines, which are for the purpose of this paper defined as the 20~K and 47~K isotherms in the disc midplane. In Fig.~\ref{coratio} we show how the radial extent of the C/O=1 region\footnote{Throughout the paper, C/O is intended as the gas-phase C/O abundance}, as well as its location, change with gas loss. Both this region and the region of C/O=0.8 between the H$_2$O and CO$_2$ snowlines shrink drastically and migrate closer to the star as a result of gas loss. 

We have shown that the CO$_2$ snowline follows an inward migration analogous to that of the CO snowline, with the different processes we investigated, such as gas loss and dust growth. An exception is the clearing of the inner disc, where we witness enhanced heating in the regions behind the inner disc edge, as shown in Fig.~\ref{rin}. The CO$_2$ snowline is pushed away from the star when $R_{in}$ exceeds 10~au in our model. For different stellar properties and especially in case of T Tauri stars this may happen at considerably smaller hole sizes.

The heating by irradiation from the star, in absence of accretion heating, is challenging to compute in the first few au due to the high optical depth of these regions. The results of our models are therefore not applicable to the regions of interest for the H$_2$O snowline. These regions are explored in \citet{min2011}, for example.

\section{Observational implications}\label{obs}

\subsection{Spectral energy distribution}\label{sed}

\begin{figure*}
 \includegraphics[width=17.5cm]{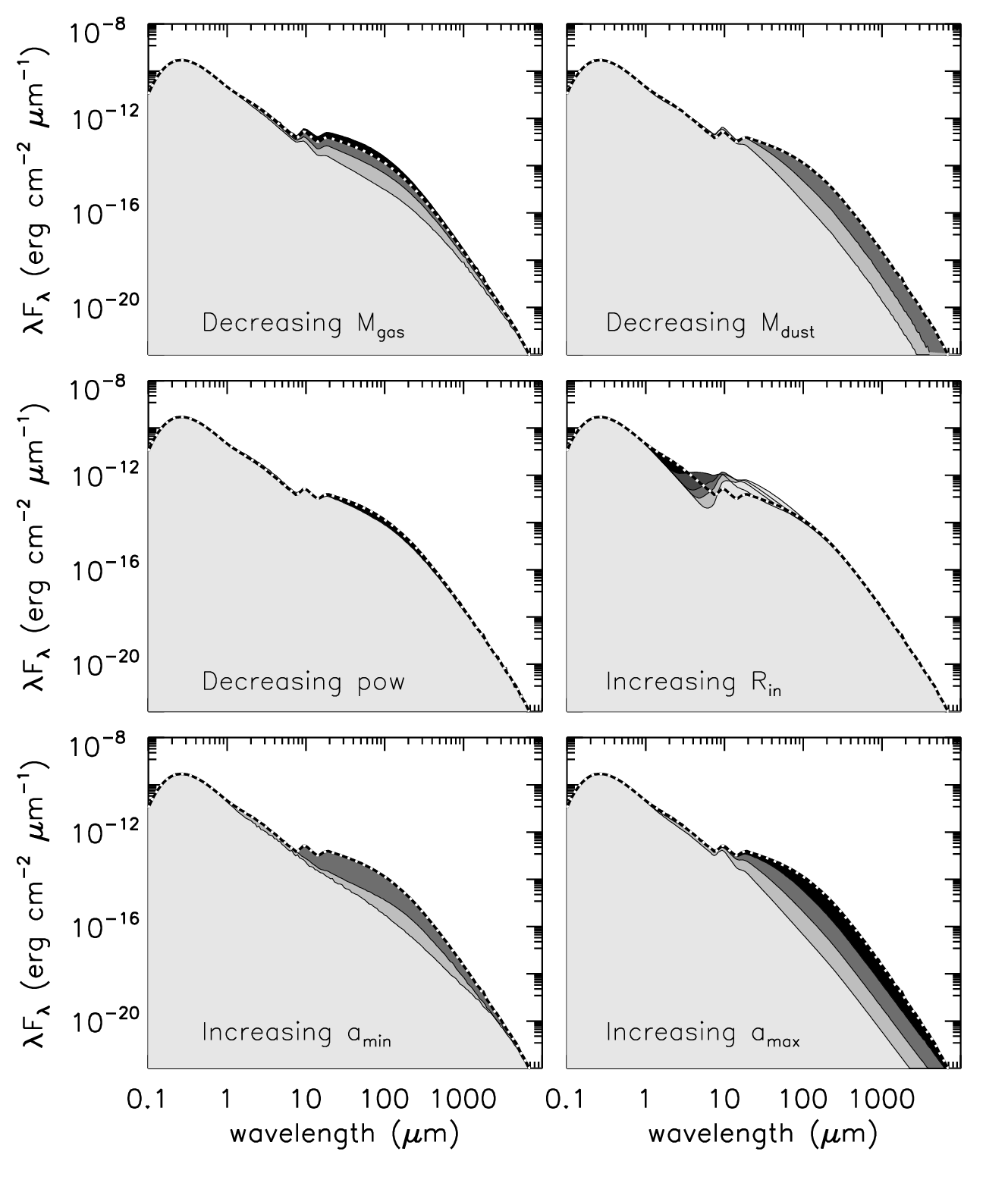}
 \caption{Spectral energy distributions for different parameter variations. The fiducial model is shown with the dotted black an white line. For each parameter variation the SEDs are shown with progressively lighter shade of grey. Tab.~1. shows the values of the varied parameters. For simplicity, in the case of R$_{in}$ we show only a subset of models, with R$_{in}$=0.24,2.5,7.5,12.5,25~au. }
\label{seds}
\end{figure*}

The presence of discs is most readily identified through the infrared excess emission of the circumstellar dust and multiwavelength photometry is available for almost all discs. For this reason, analysis of the SED is the most common method to study disc structure. Such studies often focused on physical properties of the dust, such as size, composition, spatial distribution and total mass, in order to explain the observed differences in SEDs of discs. One of the most widely invoked interpretations of the SED diversity is grain growth, shown to significantly affect the level of disc excess emission due to settling \citep{DD2004a, meijer2008}. To add to these existing studies, we model gas and dust evolution processes separately, meaning that we change the gas mass separately and independently of other disc parameters to investigate its influence on the disc structure and SED. On the other hand, we investigate the effects of dust loss and grain growth while keeping the gas mass fixed. Similar approach has been used, for example, in \citet{woitke2016} with a significant difference that in their models disc scale height is prescribed and not computed. Therefore the heating, settling and hydrostatic equilibrium are not consistent, like in our models. In addition, we investigate the effects of gas mass independently.

In Fig.~\ref{seds} we give an overview of the effects that different parameters we varied have on the SED. We assume a distance of 100~pc, and a viewing angle of 45$\degr$ from face-on. Models of gas loss, with decreasing $M_{gas}$, clearly affect the SED as significantly as dust loss, or grain growth processes. The dust excess emission at all wavelengths decreases when gas is lost because the gas provides the pressure to the smallest dust particles following $H_{dust} \propto H_{gas} \sqrt{\alpha\rho_{gas}}$ (see Sect.~\ref{fiducial}). As discussed in Sect.~\ref{results}, this makes the gas, alongside the size of the smallest dust grains, a key factor in setting the vertical thickness and heating of the disc, which in turn determines the level of the thermal emission. 

It is clear from Fig.~\ref{seds} that some processes have generally similar effects on the SED, e.g. gas loss and removal of the smallest grains, or dust loss and growth of the largest grains. This means that the SED analysis alone would not suffice to discern between different processes causing these effects and complementary data are needed to constrain them individually. Some of the effects of different parameters on the SEDs and parameter degeneracies are already known \citep{meijer2008,woitke2016} but the relation between the excess emission and disc temperature has not been fully explored as often disc scale height was prescribed rather than computed. The four aforementioned processes all have decreasing trends of excess emission from the IR to millimetre wavelengths (upper two and bottom two panels of Fig.~\ref{seds}). It is tempting to associate these excesses with disc temperature. However, the corresponding trends in $R_{CO}$ are diverse(see Fig.~\ref{snowy}). There is a large decrease in $R_{CO}$ for all model variations except for the $a_{max}$ group of models where only a minor decrease in $R_{CO}$, i.e., disc temperature, is found. Interestingly, some parameters affect the temperature but not the SED, and vice-versa. For example, the SEDs of the $pow$ series of models appear to be the same, but the temperature structures differ. On the other hand, $a_{max}$ affects the SED notably, while the temperature structure in these models remains almost unchanged. 
These results lead us to conclude that SED analysis and modelling do not uniquely determine the disc temperature structure. A practical example of this is given in our paper \citet{boneberg2016}, where we find different midplane temperatures in a number of different physical models that fit the SED of HD~163296. This stresses the importance of measuring disc midplane temperature directly, rather than relying on SED models. Possible observational approaches are discussed in Sect.~\ref{lines}.

\begin{figure}
 \includegraphics[width=9.cm]{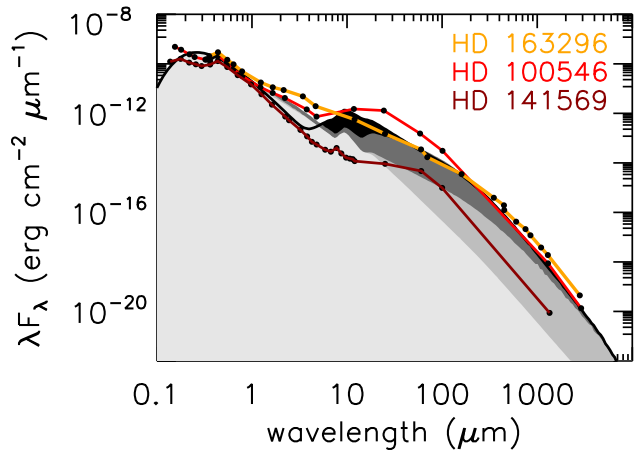}
 \caption{Spectral energy distributions, scaled to 100~pc, of three selected protoplanetary discs around A-type stars: HD~163296 (orange), HD~100546 (red) and HD~141569 (brown). For comparison of the overall levels of excess, several models are shown: $R_{in}$=7.5~au (black), the fiducial model (dark grey), $M_{gas}$=$10^{-4}$~M$_{\odot}$ (grey), and $a_{max}$=1~km (light grey).}
\label{realseds}
\end{figure}

In Fig.~\ref{realseds} observed SEDs are shown for three well studied Herbig Ae stars, hosting discs of very different properties: HD~100546 has a disc with a large gap and is suspected of hosting planets \citep[e.g.][]{quanz2015}, HD~163296 has an SED that can be modelled by a continuous (non-transitional) disc \citep[e.g.][]{tilling2012}, and HD~141569 has a disc which is transitioning into the debris-disc regime judging by its low gas mass as seen in CO imaging \citep{flaherty2016}. The three discs occupy very distinct regions of the IR excess/colour diagram \citep{vanboekel2005,meijer2008} and roughly cover the range in the observed diversity of disc SEDs in Herbig Ae stars, from Group I HD~100546 to Group II HD~163296. In comparison, the grey shaded models cover similar ranges of excess emission and this goes to show that our parameter variations are similar to the SED types and excess levels seen in Herbig Ae stars. Therefore the physical regimes explored here are relevant for what is considered to be the optically thick stage of a protoplanetary disc, evolved enough to not have envelope emission, but not as evolved as the optically thin debris discs around main sequence stars. 
One of our key results is that discs can cool and settle without invoking dust evolution at all. Settling of dust and of the overall disc vertical structure can be achieved entirely through gas loss, even if the dust remains pristine throughout this process. Gas loss is an equally plausible, alternative mechanism to grain growth, and can visibly modify the SED of a disc from that of Group I to that of Group II. Therefore, Group II sources may be the gas-depleted counterparts of the Group I sources, and do not necessarily have more evolved dust. This trend could be tested through C$^{18}$O and $^{13}$CO observations of midplane gas density (see Sect.~\ref{lines}), but care must be taken to account for the possibility of gas-poor discs coming back to Group II by opening large holes.

\subsection{CO snowline observations}\label{lines}

One of the main results of our parameter study is that the location of the CO snowline, along with the snowlines of other low-temperature desorbing species, is strongly affected by the gas mass and by the size of the smallest dust particles present in the disc surface. Removal of the gas and of the sub-micron dust particles significantly decreases the efficiency with which the disc midplane is heated by the stellar irradiation. Heating is not the only factor determining the snowline location, as local vapour pressure and ice composition may affect the desorption temperature and hence the location at which desorption occurs, but its effects on the snowline location are vastly dominant over the minor effect of pressure for highly volatile species such as CO \citep{davis2005}. The CO snowline location is not determined precisely when derived solely from an SED model of a given disc, due to parameter degeneracies that plague such approach \citep{boneberg2016}. This warrants direct observations of CO snowline location, as well as complementary measurements of the midplane temperature. Measurements have recently been done with ALMA using N$_2$H$^+$ line emission, finding the snowline locations of 30~au in TW~Hya and 90~au in HD~163296 \citep[][ respectively]{qi2013a,qi2015}. This opens a promising avenue to study and spatially resolve the CO snowline, especially in Herbig~Ae discs where the snowline is located at comparably larger distances from the star than in the case of discs around fainter, T~Tauri stars. 

An alternative observational approach is to employ high-sensitivity observations of C$^{18}$O to detect the radial drop in C$^{18}$O density caused by freeze-out at the CO snowline. This more direct approach has validated the results based on N$_2$H$^+$ observations of HD~163296 \citep{qi2015}. Furthermore, observational constraints on the CO snowline location are necessary because the desorption temperature can vary and gaseous CO at temperatures lower than 20~K has been observed in some discs \citep[e.g.,][]{dartois2003,hersant2009}. The observed C$^{18}$O density gradients are set vastly by the geometry of the snow-surface layer through the disc atmosphere, and of the angular resolution achieved. In very settled discs, for example, the snow-surface layer lies closer to the midplane than in the flared disc case, and this will affect the gradient in the observed radial abundance of gas-phase C$^{18}$O. For this reason, interpretation of C$^{18}$O emission from regions beyond the snowline requires self-consistent modelling of the gas and dust disc structure using multiple observational constraints. On the other hand, spatially resolved observations of C$^{18}$O within the CO snowline are much more robust with respect to the uncertainties in disc structure and temperature. The reason for this is that the C$^{18}$O emission in the unfrozen part of the disc inwards of the CO snowline is vastly dominated by the gas density while being relatively insensitive to the midplane temperature.
Such observations begin to assess the degree of symmetry and the physical conditions of the midplane gas, greatly aided by the fact that C$^{18}$O emission is mostly optically thin.

\section{Implications for planet formation and disc evolution}\label{planetformation}

\subsection{Protoplanetary disc evolution}

An important result of our study is that significant inward migration of the CO snowline can be achieved though \textit{gas dispersal} alone, even if there is no dust evolution. In Sect.~\ref{results} we identified the gas mass, or more precisely - the gas column density, and the size of the smallest dust particles present in the disc surface as two key factors in determining the efficiency with which the disc is able to absorb the stellar emission and thereby heat the disc interior. Higher amounts of gas and smaller dust lead to more flared disc geometries and warmer disc interiors, while the loss of gas and of small dust lead to dust settling. 
Interestingly, grain growth is not the only way to remove smallest dust grains. Photoevaporative winds driven from the disc surface entrain sub-micron sized dust as well as gas \citep{takeuchi2005,owen2011}. Our findings provide an important new link between the observable properties of the disc (as discussed in Sect.~\ref{obs}) and the degree of disc evolution, in particular the gas mass, which has so far been challenging to tackle observationally. The first application of this method is shown in \citet{boneberg2016}, where we investigate the midplane gas of HD~163296 by combining the SED, CO snowline imaging and C$^{18}$O gas imaging in the context of physically consistent disc models.

We considered \textit{grain growth} in three different ways, focusing on: the growth of sub-micron sized particles (dominating effective dust surface and dust optical properties), the growth of grains, pebbles and planetesimals (containing the bulk of the dust mass) and an overall decrease of dust mass. Growth of only the large dust (beyond 1mm) does cause these large dust particles to settle but our models show that what happens at the high end of the size distribution has no significant effect on disc heating because large dust particles are merely the passive receptors of re-radiated emission in the disc interior. We have shown that only the growth of sub-micron particles makes an impact on disc temperature, and this highlights the importance of the opacity provided by the small dust in the disc surface. If this dust can grow efficiently then grain growth can lead to an increased settling of the dust and lower midplane temperatures. Absence of sub-micron grains may be detected in particularly bright and extended discs, by measuring the colour of the scattered light. For example, \citet{mulders2013} find the minimum size of grains in the surface of the disc around HD~100546 to be approximately 2.5~$\mu$m. In the context of our results, we could expect that the overall vertical structure of HD~100546 is relatively settled. The explanation for the Group I classification of this disc would then be the large mid-infrared excess caused by the disc wall at 10~au \citep{panic2014}. 

Our results show that the main processes that drive disc evolution - gas dispersal and grain growth - both drive the CO snowline to migrate inwards in our models. Although we investigate models of a bright Herbig Ae star, the same disc physics applies to the fainter T Tauri stars and the trends we identify are qualitatively the same. However, if we wish to build a complete thermal (i.e., chemical) history of the disc midplane, then we must consider the effect of the other major player - the star itself. During the first 5-10~Myr of the pre-main sequence evolution, luminosities of low-mass T Tauri stars decrease, while the opposite happens for the radiative intermediate-mass stars, which increase drastically in luminosity and effective temperature along their pre-main sequence tracks \citep[e.g.,][]{siess2000}. Without sophisticated modelling, we can conclude that for the low-mass stars the midplane temperatures decrease during the crucial stages of giant planet formation (the gas-rich stage). For these stars, both disc evolution and stellar evolution drive the temperature down with time, settle the discs to flatter configurations and bring the CO snowlines closer to the star. For intermediate-mass stars the situation is clearly more complex, due to the opposing effects of the stellar and disc evolution on the disc heating. Any result about the trend in evolution of the midplane temperature with age, in case of intermediate-mass stars (Herbig Ae) requires an assumption about the gas dispersal timescale. This timescale is likely to be different for the intermediate-mass stars because EUV and FUV dispersal may play a major role there, unlike for the X-ray dispersed T Tauri stars. A conclusion we can draw based on these considerations is that there are marked differences in the disc temperature (and therefore settling) trends during giant planet formation phase for low-mass and intermediate mass stars. The former discs settle steadily. This enhances the density of solids throughout the disc midplane, and moves their CO snowlines towards the regions of higher densities of solids, closer to the star. It is possible to test this hypothesis observationally, for example, by employing CO line ratios to compare the temperature of the disc interior in samples of stars of different ages, or by measuring their CO snowline locations.
Discs around intermediate-mass stars may maintain their snowlines at relatively large radii for a longer time due to the competing effects of disc and stellar evolution. These differing trends are intriguing because the division happens near 1.5~M$_{\sun}$, in striking similarity with the stellar mass at which differences are seen in hot Jupiter populations around exoplanets. It would be extremely interesting to provide better observational constraints regarding disc evolution in the 1-3~M$_{\sun}$ regime, extending the current statistical studies to include the intermediate-mass stars wherever possible.

 \subsection{Disc chemistry and (exo-)planet compositions}

We find that for a range of disc properties, the CO snowline may be located a few tens or a few hundred au from the star, for a typical Herbig Ae star. A difference of one order of magnitude in dust density and vastly different orbital timescales between these two extremes may lead to different outcomes of planet formation processes. From this, we can conclude that a planet whose composition suggests formation at the CO snowline may have formed through different mechanisms, depending on where the CO snowline was located at the time the planet was formed. This is particularly interesting in the view of the results of \citet{ali-dib2014}, who suggest that Uranus and Neptune may have formed at the CO snowline in the Solar protoplanetary disc, judging by their chemical compositions.

 Even though CO snowline does not persist at an invariable distance from the star, it is important to search for evidence of planets being assembled at the CO snowline. Such evidence is most likely to be indirect, given the challenges involved in detecting planets directly in protoplanetary discs. Recent studies of HD~163296 are beginning to shape the observational picture of this process, with a gap and a ring derived from ALMA images in the regions close to the observed CO snowline location \citep{zhang2016,guidi2016} as well as evidence from polarimetric imaging \citep{garufi2014}.

We find intriguing differences between low-mass and intermediate-mass stars ($<$1.5~M$_{\sun}$ vs 1.5-3.0~M$_{\sun}$), in that the discs of the former continuously decrease in temperature, while the latter have opposing mechanisms that may maintain the snowline at larger distances from the star even at later stages of disc evolution (roughly 10~Myr, corresponding to the age of the oldest observed gas-rich protoplanetary discs). This means that the region of high gas phase C/O abundance (C/O=1) in these two types of discs is very different, continuously small and shrinking in the case of T Tauri stars and comparably larger and possibly more static in case of Herbig Ae stars. It will be interesting to identify any difference in the chemical compositions of exoplanet populations between the stellar mass regimes below and above 1.5~M$_{\sun}$), and try to link this to the differing disc midplane chemistry.

\section{Conclusions and future prospects}\label{conclusions}

To summarise, this paper uses state-of-the-art modelling tools to investigate the impact of different disc evolution processes on the heating of the disc interior, in particular the disc midplane, where planets form and evolve. A very important aspect of our method is to give full consideration to the role of gas pressure and the balance between the pressure support by which the gas lifts dust particles, the heating that these particles are able to provide in turn, by absorbing the starlight in the disc surface, and the feedback of the re-radiated emission on the disc temperature and the hydrostatic pressure. This close relation between the gas and the dust leads us to focus on both gas and dust evolution and calculate the temperature of the dense disc interior in a fully self-consistent manner. We identify the key disc properties related to heating and investigate the observable impact these parameters have on disc midplane properties: the location of the CO snowline and the extent of the C/O=1 region in the gas. While our fiducial model is that of a typical Herbig Ae star, the qualitative trends found here apply to discs around both low-mass and intermediate-mass stars alike, because the processes of absorption/re-emission of radiation, temperature and pressure balance in their discs are physically identical.

Gas loss and disc heating
We have demonstrated that gas dispersal has an observable impact on the temperature and flaring of protoplanetary discs, leading to lower temperatures in the disc interior and dust settling. The latter can be directly measured through CO snowline imaging and SED modelling. As a result, settled disc geometries do not directly imply a higher degree of grain growth. In fact, the only other disc property besides the gas density that determines the disc temperature and settling is the size of the smallest dust particles. Such particles can be removed from the disc surface by photoevaporative winds, for example.

CO snowline location
The CO snowline is closely tied with the temperature, such that the above considerations extend to the issue of where the CO snowline is located in discs and the breadth of the region of C/O=1 gas phase abundance between the CO and CO$_2$ snowlines. Therefore, the loss of gas and small dust, for example in a photoevaporative flow, results in a CO snowline location considerably closer to the star, with respect to where it would be if the disc were gas rich. For a typical Herbig Ae star, we find differences in the CO snowline location that are larger than 100~au, so the snowline could be anywhere between few tens of au from the star in a gas-poor disc to around 200~au in a gas-rich disc.

Evolutionary trends for discs and implications for exoplanets
The conclusions above were made based on an assumption that the stellar radiation field does not vary. While it is difficult to incorporate both stellar and disc evolution effects in a single model, due to uncertain gas dispersal timescales, we can comment on general trends.
For low-mass stars, luminosity decreases during the pre-main sequence and this, alongside disc evolution, additionally favours disc settling with age. Therefore, disc and stellar evolution both lead to inevitably colder and more settled discs. This could be put to a test by observing CO snowline locations in discs around T Tauri stars with different ages.
For intermediate-mass stars, the increasing stellar luminosity and effective temperature on the pre-main sequence tracks are able to counterbalance the `settling' effect of photoevaporation. The differing evolution trends we infer for low- and intermediate-mass stars concerning the outer regions where gas giants form are very intriguing because this division happens around 1.5~M$_{\sun}$, the same stellar mass above which hot Jupiters become scarce, and over which the frequency of giant planets steadily increases. Another interesting aspect is that the potentially gravitationally unstable, gas rich discs with pristine dust around a very young pre-main sequence intermediate mass stars are warmer and have a higher C/O ratio in the gas of the outer disc than their counterparts around low-mass stars, so we may expect to see some evidence of differentiation in the chemical composition of gas giants around intermediate mass stars with respect to those around low-mass stars. This may be addressed with future instruments JWST and E-ELT.

The evolution of the Herbig Ae population
We show that opening an inner hole in the disc can cause the disc to display an apparently flared SED, while the bulk temperature of the disc interior or its flaring over larger scales does not change. The hole opening enhances disc heating in a limited radial region directly behind the disc wall, and it is this region that causes a decrease in the near-IR and an increase in mid-IR and/or far-IR excesses. The temperature and emission from the remainder of the disc, which contains the bulk of the disc mass if the disc is much larger than the hole, remain unchanged. Hole opening is therefore an effective mechanism for a disc around a Herbig star to `migrate' from Group II to Group I classification, in the opposite direction than dust evolution. This is in line with the observational findings of \citet{menu2015}. Because of this, we should expect the Group I (commonly identified as `flared') to contain a mix of more and less evolved discs. This happens only once the hole is larger than $\approx10$~au, so it may be possible to observationally identify these `evolved' stragglers in the Group I sample \citep{maaskant2013}. A long standing hypothesis has been that disc settling is caused by dust evolution. By studying the role of gas in disc heating we identify an alternative method to settle discs, and for Group I sources to 
evolve into Group II, is by gas loss. This method can work effectively even if the dust remains pristine throughout the stage of gas-loss, and does not require any dust evolution. It will therefore be interesting to see, e.g. with ALMA, whether Group II discs are gas-poor compared to the Group I population.

\section*{Acknowledgments}

The work of O.P. is supported by the Royal Society Dorothy Hodgkin Fellowship. Work of O.P. presented here was also supported by the European Union through ERC grant number 279973. The authors thank Inga Kamp, Carsten Dominik, Cornelis Dullemond, Cathie Clarke, Mark Wyatt and Uma Gorti for useful discussions.



\bibliographystyle{apj}
\bibliography{new}
\label{lastpage}




\bsp	
\label{lastpage}
\end{document}